# Revealing the Topology invariance of vectorial vortex beam in complex media

SHUAILING WANG,[1] JINGPING XU[1] AND YAPING YANG[1]

[1]*School of Physics, University of the Witwatersrand, Johannesburg 2000, South Africa*

**Abstract:**

Orbital angular momentum (OAM), a topological degree of freedom of light, is theoretically invariant under continuous deformations; yet, its physical observability degrades precipitously in complex media, creating a fundamental "topology-observability gap." Here, we propose a novel paradigm for topological measurement based on the non-separable coupling between polarization and topological features in vectorial vortex beam. By constructing a topological non-separability measure $\Xi_\ell$ derived from global Stokes fields, and integrating it with a physics-guided machine learning calibration framework that combines Bayesian Gaussian process regression with XGBoost-driven adaptive model selection, we achieve high-fidelity identification of topological features up to $\ell = 200$. Crucially, this robustness persists even when beam intensity and phase structures are completely distorted by extreme complex media, including strong atmospheric turbulence, oceanic turbulence, and high-temperature jet exhausts. This approach overcomes the dual bottlenecks of limited accessible OAM modes and susceptibility to perturbations that constrain conventional methods. Our work not only bridges the fundamental divide between topological theory and physical observability, but also establishes a robust framework for the reliable deployment of high-dimensional OAM in real-world complex environments, promising exciting advancements for wireless optical communications, remote topological sensing, and classical analogs of quantum information protocols.

## 1. Introduction

Vortex beams, light fields that carry orbital angular momentum (OAM), have attracted broad interest since their inception[1, 2]. This distinctive attribute has catalyzed breakthroughs across diverse domains, including optical tweezers[3-5], wireless optical communications [6-8], super-resolution microscopy[9-11], and remote sensing[12-14], thereby significantly advancing modern photonics. However, the very OAM that endows vortex beams with such rich functionality also renders them particularly susceptible to degradation in complex propagation media[15-17]. This susceptibility constitutes a major bottleneck limiting the broader adoption of vortex beams across various optical domains. Consequently, how to effectively suppress or overcome the detrimental effects of complex propagation media on vortex beams has long remained a critical and outstanding challenge in the field[18].

Recent advances have unveiled a promising paradigm shift: vectorial vortex beams exhibit remarkable resilience against channel-induced distortions. Crucially, even when severe turbulence or scattering catastrophically scrambles the spatial intensity and phase profiles of a beam, the underlying polarization nonuniformity of vector fields remains remarkably preserved—provided one adopts an appropriate measurement framework[19-21]. Specifically, by abandoning rigid reliance on the original modal basis and instead constructing a channel-adapted analysis basis via Stokes polarimetry, the robustness of vectorial vortex beams emerges not from morphological stability, but from the invariance of its polarization texture[19]. This insight reframes the problem: rather than attempting to suppress disturbances, one can actively harness them—by flexibly tailoring the detection basis to match the perturbed channel, thereby efficiently decoding information. This conceptual pivot not only deepens our understanding of vortex beams propagation in complex media, but has also opened up practical new pathways toward enhancing the utility of vortex beams in real-world environments.

As a mathematical branch that studies properties invariant under continuous deformation, topology reveals that even when an object undergoes drastic geometric distortion and significant changes in appearance, its intrinsic topological essence remains intact. This provides a profound perspective for understanding the OAM of vortex beams: the OAM carried by vortex beams of different orders can be regarded as a topological feature. This topological feature endows vortex beams with an inherent stability, suggesting that—at least in theory—a measurement scheme could be devised to effectively distinguish and identify them via their topological feature, even when their wavefronts suffer continuous distortions. However, extensive studies have shown that in complex media, such as atmospheric or oceanic turbulence, the OAM spectrum of vortex beams often exhibits significant dispersion[22, 23]. This manifests as misinterpretation of the original topological feature at the receiver[24], reduced received probability[25, 26], and enhanced crosstalk between topological feature[8, 27, 28]. This indicates that although the topological feature itself is a mathematical invariant, conventional measurement approaches based on scalar-mode projection are highly susceptible to channel perturbations, leading to a " topology-observability Gap."

To address this contradiction, we propose a novel topological feature measurement framework. Rather than relying on fixed spatial mode bases, our approach exploits the non-separable correlation between polarization and OAM degrees of freedom in vectorial vortex beams. By integrating global polarization descriptors, we construct a topological fingerprint that remains insensitive to channel perturbations. Owing to the intrinsic invariance of topological features in vortex beams, their underlying topological feature information can be accurately retrieved via this adapted criterion, even when the beam's spatial morphology is severely distorted. Consequently, our method achieves robust preservation of the topological feature. It not only respects the topological feature mathematically, but also ensures anti-interference capability at the level of physical measurement. This innovative framework not only reinterprets the topological stability of vortex beams in perturbed environments, but also provides a tool of both theoretical depth and practical utility for high-reliability OAM-multiplexed communications, perturbation-resistant optical sensing, and precision sensing applications.

## 2. Bridging the Topology-Observability Gap with Non-Separable Correlations

OAM, as an intrinsic topological degree of freedom of vortex beam, holds immense potential for information carrying capacity owing to its high-dimensional orthogonal mode space [29]. However, in the three decades since its discovery, a fundamental bottleneck has persistently hindered its practical deployment: although the topological feature $\ell$ is a strict invariant under continuous deformation in a mathematical sense, its physical observability becomes exceptionally fragile in complex media. This fragility originates from two coupled physical bottlenecks. First, the transverse scale of high-order vortex beams expands with $|\ell|$, leading to severe truncation in finite-aperture systems. Second, random refractive index fluctuations in complex media disrupt the helical phase structure, causing nonlocal spectral diffusion of the OAM modes and rendering high-order topological feature difficult to reliably identify at the receiver. Through customized optical field modulation and a high-precision OAM detection algorithm, we resolve the conflict between beam expansion and aperture constraints [30]. Consequently, we successfully elevate the maximum order of stably usable topological feature in free space from 20 to 100.

However, practical transmission environments are not always free space; wireless optical links must contend with typical complex media such as atmospheric turbulence, oceanic turbulence, and jet engine turbulence. Turbulence-induced phase randomization leads to severe degradation of vortex beam performance [31-34]. As illustrated in Fig. 1, conventional algorithms can reliably identify the topological feature of vortex beam in free space with 100% fidelity.

However, the introduction of turbulence induces significant decoding errors, with this degradation becoming markedly more severe as turbulence strength increases (i.e., as the dissipation of kinetic energy per unit mass of fluid $\varepsilon$ decreases). Under weak turbulence conditions ($\varepsilon = 10^{-5}\,\mathrm{m^2/s^3}$), the system maintains relatively modest errors, supporting robust OAM information transmission. However, under strong turbulence ($\varepsilon = 10^{-9}\,\mathrm{m^2/s^3}$), turbulence eddies severely distort the helical phase structure of the vortex beam, inducing substantial mode crosstalk that renders the detected topological feature highly distorted and fluctuating, rendering them unreliable for information encoding or decoding. This fragility stems from the scalar nature of conventional OAM measurements, the information relies entirely on spatial phase structures that lack topological protection against perturbations in the complex media. This reveals a profound paradox: the mathematical invariance of topological charge does not automatically guarantee its physical observability. This "topology-observability gap" represents not merely a technical bottleneck, but a fundamental barrier bridging abstract topological theory and practical information transmission, one that critically constrains the transition of OAM technologies from controlled laboratories to dynamic, unpredictable real-world environments.

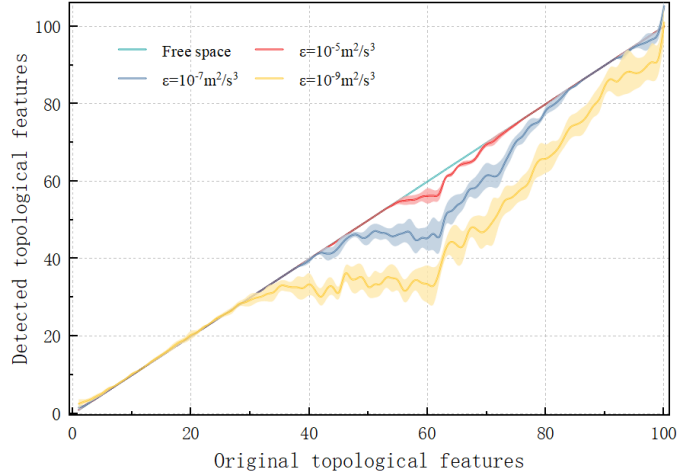

**Fig. 1.** The relationship between the detected topological features of vortex beams and their original topological features under the influence of turbulence characterized by different $\varepsilon$. Solid lines represent the mean values of the detection results, while shaded regions indicate the corresponding standard deviations.

To bridge this bottleneck—topology-observability gap, we invoke the principle of topological observability based on non-separable degrees of freedom. The core of this approach lies in the non-separable correlation between polarization and topological feature. Accordingly, we first construct vectorial vortex beams of the following form:

$$|\Psi\rangle = |e_1\rangle|o_1\rangle + |e_2\rangle|o_2\rangle \tag{1}$$

Here, the polarization degree of freedom is represented by a pair of orthogonal basis vectors $|e_1\rangle = |H\rangle$ and $|e_2\rangle = |V\rangle$, while the topological feature degree of freedom is given by the orthogonal basis modes $|o_1\rangle = \ell$ and $|o_2\rangle = \ell + N$, corresponding to distinct OAM states. Crucially, these two subspaces are not independent: the polarization state at each spatial point is inextricably linked to its topological feature, forming an inseparable composite structure. This non-separability ensures the topological feature from an isolated phase singularity to an entity embedded within a vectorial topological complex. In this framework, physical observables are dictated by the global geometric correlation between polarization and phase, rather than relying on local intensity distributions or phase gradients. Within this framework, the vectorial vortex beams can be rigorously described as a non-separable classical state in the composite Hilbert space $\mathcal{H}_{\text{pol}} \otimes \mathcal{H}_{\text{OAM}}$. The strong correlation structure between its degrees of freedom is mathematically isomorphic to quantum entangled states, thereby inheriting the latter's robustness against local perturbations. It is precisely this correlation, protected by the intrinsic symmetry of the structure, that provides the physical basis for achieving observable invariance of topological information in complex media.

Upon propagation through complex media satisfying unilateral unitarity, although the OAM modes carried by individual polarization components disperse into a high-dimensional Hilbert space, their relative geometric phase structure and modal orthogonality are preserved due to the unitarity symmetry of the channel. This intrinsic invariance implies the existence of an observable global order parameter capable of quantitatively characterizing the strength of non-separable correlations between polarization and topological degrees of freedom. Inspired by the concurrence metric in quantum information theory, we construct a topological non-separability measure to quantify the topological robustness of cross-degree-of-freedom correlations in this vectorial vortex beams:

$$\Xi_\ell = \sqrt{\frac{(S_0 - S_1)^2 + S_2^2 + S_3^2}{(S_0 + S_1)^2}} \quad (2)$$

The topological non-separability depends on the global Stokes field and can be understood as a weighted deviation of the polarization distribution from the equatorial plane (linear polarization) on the Poincaré sphere. Crucially, its value embeds a topological feature uniquely characteristic of the OAM order. Benefiting from its strict invariance under unilateral perturbations, this measure constitutes a channel-robust topological feature, enabling the phase-free and anti-interference readout of high-dimensional OAM states via polarization-resolved intensity imaging alone. Fig. 2 intuitively illustrates this principle: even when the beam wavefront is severely distorted by strong turbulence, the topological non-separability retains its distinctive fingerprint in one-to-one correspondence with $\ell$, thereby supporting highly robust topological identification. However, Fig. 2 also reveals a critical challenge: the mapping between topological non-separability and $\ell$ is highly nonlinear, with measurement values spanning hundreds of orders of magnitude from low-order to high-order topological features. Under this condition, directly inverting $\ell$ from the raw observable becomes practically infeasible; minute measurement noise or systematic deviations can trigger erroneous integer topological feature, particularly in the high-order where this sensitivity becomes most pronounced.

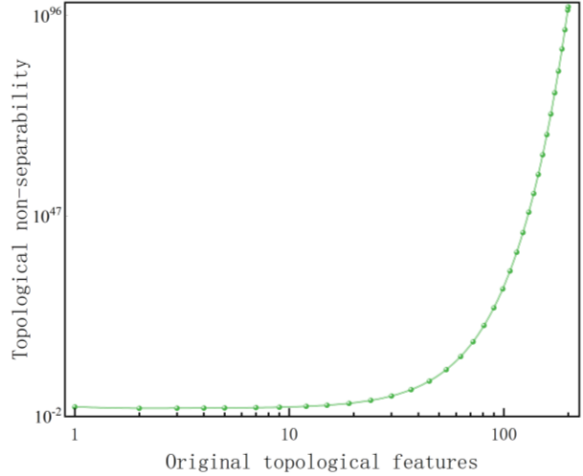

**Fig. 2.** Evolution of topological non-separability as a function of original topological features.

## 3. From Nonlinear Observable to Linear Encoding via Physics-Guided Learning

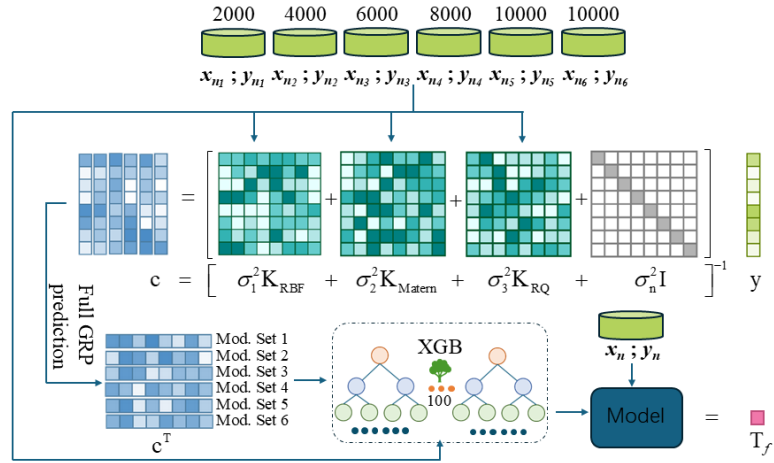

**Fig. 3.** Two-stage framework for topological fingerprint calibration. Six GPR models (five partitioned, one global) are trained with physics-guided kernels. An XGBoost classifier selects the optimal model per input based on features, enabling high-fidelity mapping from topological non-separability to topological fingerprint $T_f$.

To achieve linearly resolvable encoding of high-order topological features, we construct a physics-guided topological features calibration mapping model that precisely translates the highly nonlinear topological non-separability onto the integer topological fingerprint $T_f$. As illustrated in Fig. 3, we propose a two-stage hybrid modeling framework that integrates Bayesian Gaussian process regression (GPR) with XGBoost-driven adaptive model selection. First, based on the theory of vectorial vortex beams, we generate 100,000 data samples spanning a broad parameter space encompassing varying ring thicknesses, ring radius, and topological features $\ell \in [0, 200]$. To construct an ensemble of expert models with complementary characteristics, we build six model sets with different training sample sizes: 2000, 4000, 6000, 8000, 10,000, and 10,000 samples, respectively. Each dataset is randomly partitioned into training (80%) and testing (20%) sets. For the first five model sets, we adopt a partitioned training strategy, constructing independent GPR models within each subdomain delineated by the partition nodes $\ell = 50, 100, 140, 170$ ; for the sixth model sets, a single global model is trained on the entire dataset. During the training of each model set, we construct physics-inspired composite kernel functions, including the Radial Basis Function (RBF) kernel, Matérn kernel, Rational Quadratic (RQ) kernel, and White Noise kernel. To mitigate local optima issues, each model undergoes five random restarts during training, and the model achieving the highest proportion of test samples with absolute error below 0.001 is selected as the optimal representative model for that subdomain.

In the second stage, to achieve sample-level adaptive optimal selection among the six candidate expert models, we engineered an XGBoost classifier. This classifier dynamically identifies the optimal expert model for each input by learning a nonlinear mapping from a feature vector (comprising original physical parameters, predictions from all six expert model, and GPR uncertainty metrics) to the index of the best-performing model. Through supervised learning, the XGBoost model learns a nonlinear mapping from the feature space to the optimal model index, enabling adaptive model selection for different input samples. Ultimately, this framework successfully achieves high-fidelity, robust mapping from the highly nonlinear, near-100-order-of-magnitude spanning topological non-separability to the topological fingerprint $T_f$. This provides a scalable physics-machine learning synergistic solution for reliable readout of high-dimensional OAM information encoding in complex media.

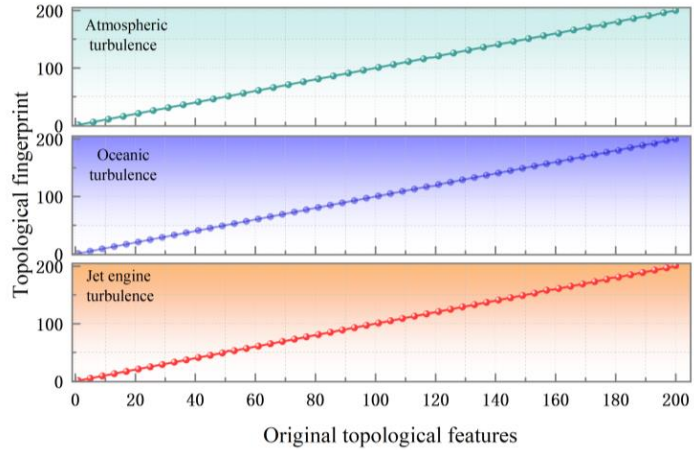

**Fig. 4.** Universal robustness validation of the topological non-separability measurement paradigm in extreme complex media.

As illustrated in Fig. 4, we systematically validated the universal robustness of this framework across three representative classes of extreme complex media: strong atmospheric turbulence ( $C_n^2 = 10^{-13} \, \text{m}^{-2/3}$ ), dynamic oceanic turbulence (double-diffusive salt-temperature instabilities), and high-temperature jet engine exhaust plumes (heterogeneous thermal perturbations at varying incident mode). The results demonstrate that the vectorial vortex beams measurement paradigm, founded on topological non-separability, yields highly consistent topological feature identification across these markedly different complex media. Under all tested scenarios, the topological feature identification accuracy remained stably above 95%. This performance significantly surpasses conventional scalar topological features detection schemes based on scalar modal projection, extending the usable range of topological features for vortex beams from the traditional limit of $\ell \approx 20$ to $\ell \approx 200$. Notably, even when the intensity distribution of the beam is completely distorted by complex medium-induced perturbations and the phase singularity structure becomes entirely indiscernible, the topological fingerprint remains remarkably stable, with its embedded topological feature remaining clearly distinguishable. This phenomenon reveals a fundamental shift in measurement paradigm: information has been transferred from "morphology" to "correlation." These findings confirm that our proposed framework successfully resolves the two coupled physical bottlenecks plaguing traditional vortex beams—finite aperture truncation and turbulence-induced mode diffusion. Our approach not only enables the effective utilization of high-order vortex beams with topological feature up to 200, but more critically, ensures that the OAM information carried by vortex beams can be reliably read out at the receiver with high fidelity and anti-interference capability, even under extreme perturbation conditions. This work establishes a solid physical foundation and technological pathway for the practical deployment of high-order OAM information transmission in real-world complex environments.

## 4. Conclusions

We propose a highly robust topological feature measurement framework that resolves the long-standing challenge of the "topology-observability gap"—a fundamental bottleneck hindering the practical deployment of OAM in real-world complex environments. Our key breakthrough lies in redefining the problem: rather than attempting to preserve beam morphology against perturbations, we exploit the intrinsic non-separable correlation between polarization and topological features in vectorial vortex beams to extract their inherent topological invariance. Based on this principle, we introduce a novel measurement paradigm centered on a "topological fingerprint." The resulting measure of topological non-separability maintains remarkable stability even when the beam's intensity and phase distributions are severely distorted by complex media. To overcome the analytical challenges posed by the highly nonlinear mapping between the raw observable topological non-separability and the topological feature, we developed a physics-guided machine learning framework. By synergizing Bayesian GPR with an XGBoost-driven adaptive model selection mechanism, this framework successfully maps the topological non-separability spanning over hundreds orders of magnitude in dynamic range onto a linearly resolvable topological fingerprint space. Consequently, we achieve high-fidelity identification of topological features of vortex beams under extreme perturbative environments, including strong atmospheric turbulence, oceanic turbulence, and jet engine turbulence. This paradigm shift yields two critical breakthroughs: First, it resolves the aperture-induced truncation bottleneck, significantly extending the range of reliably identifiable topological features from the traditional limit of $\ell \approx 20$ to $\ell \approx 200$. Second, it overcomes turbulence-induced modal diffusion, rendering the topological features of vortex beams virtually immune to propagation through complex media. Our work establishes a universal framework for the practical utilization of high-dimensional topological degrees of freedom in photonic information systems, marking a pivotal step in transitioning topological photonics from idealized laboratory conditions to real-world complex scenarios. This opens new avenues for perturbation-resistant optical communications, remote sensing, and classical analog of quantum information protocols.

**Funding.** National Natural Science Foundation of China (12174288, 12274326, 12174243, 61905186); National Key Research and Development Program of China (2021YFA1400602); Shanghai Aerospace Science and Technology Innovation Foundation (SAST-2022-069); Fundamental Research Funds for the Central Universities (ZYTS23078).

**Disclosures.** The authors declare no conflicts of interest.

**Data Availability.** Data underlying the results presented in this paper are not publicly available at this time but may be obtained from the authors upon reasonable request.

**References**

[1] L Allen, M W Beijersbergen, R J Spreeuw, et al. Orbital angular momentum of light and the transformation of Laguerre-Gaussian laser modes. Physical Review A, 1992, 45(11): 8185-8189.
[2] S M Barnett, L Allen. Orbital angular momentum and nonparaxial light beams. Optics Communications, 1994, 110: 670-678.
[3] M P MacDonald, L Paterson, K Volke-Sepulveda, et al. Creation and manipulation of three-dimensional optically trapped structures. Science, 2002, 296(5570): 1101-1103.
[4] V Garces-Chavez, D McGloin, H Melville, et al. Simultaneous micromanipulation in multiple planes using a self-reconstructing light beam. Nature, 2002, 419(6903): 145-147.
[5] X Z Li, H X Ma, H Zhang, et al. Is it possible to enlarge the trapping range of optical tweezers via a single beam? Applied Physics Letters, 2019, 114(8).
[6] H Chang, P Xu, H Yao, et al. Nonprobe Adaptive Compensation for Optical Wireless Communications Based on Orbital Angular Momentum. IEEE Transactions on Wireless Communications, 2024, 23(8): 9033-9043.
[7] J Wang, J-Y Yang, I M Fazal, et al. Terabit free-space data transmission employing orbital angular momentum multiplexing. Nature Photonics, 2012, 6(7): 488-496.
[8] J Liu, J Zhang, J Liu, et al. 1-Pbps orbital angular momentum fibre-optic transmission. Light: Science & Applications, 2022, 11(1): 202.
[9] F O Fahrbach, P Simon, A Rohrbach. Microscopy with self-reconstructing beams. Nature Photonics, 2010, 4(11): 780-785.
[10] L Gao, L Shao, B C Chen, et al. 3D live fluorescence imaging of cellular dynamics using Bessel beam plane illumination microscopy. Nature Protocols, 2014, 9(5): 1083-1101.
[11] T Vettenburg, H I Dalgarno, J Nylk, et al. Light-sheet microscopy using an Airy beam. Nature Methods, 2014, 11(5): 541-544.
[12] Z Chen, U Daly, A Boldin, et al. Weather sensing with structured light. Communications Physics, 2025, 8(1): 105.
[13] R Yali, W Yufeng, S Zeping, et al. Investigation of a vortex beam-based inversion method for aerosol particle size distribution. Optics Express, 2024, 32(26): 47515-47531.
[14] S Wang, Z Zhao, M Cheng, et al. Obstacle inversion based on the self-healing property of structured light. Sci Rep, 2025, 15(1): 24379.
[15] A E Willner, H Huang, Y Yan, et al. Optical communications using orbital angular momentum beams. Advances in Optics and Photonics, 2015, 7(1): 66-106.
[16] A E Willner, K Pang, H Song, et al. Orbital angular momentum of light for communications. Applied Physics Reviews, 2021, 8(4): 041312.
[17] M Cheng, W Jiang, L Guo, et al. Metrology with a twist: probing and sensing with vortex light. Light: Science & Applications, 2025, 14(1): 4.
[18] C Peters, V Cocotos, A Forbes. Structured light in atmospheric turbulence—a guide to its digital implementation: tutorial. Advances in Optics and Photonics, 2025, 17(1).
[19] I Nape, K Singh, A Klug, et al. Revealing the invariance of vectorial structured light in complex media. Nature Photonics, 2022, 16(7): 538-546.
[20] C Peters, M Cox, A Drozdov, et al. The invariance and distortion of vectorial light across a real-world free space link. Applied Physics Letters, 2023, 123(2): 021103.
[21] K Singh, I Nape, W T Buono, et al. A Robust Basis for Multi‐Bit Optical Communication with Vectorial Light. Laser & Photonics Reviews, 2023, 17(6): 2200844.
[22] J Zeng, X Liu, C Zhao, et al. Spiral spectrum of a Laguerre-Gaussian beam propagating in anisotropic non-Kolmogorov turbulent atmosphere along horizontal path. Optics Express, 2019, 27(18): 25342-25356.
[23] J Ou, Y S Jiang, J H Zhang, et al. Spreading of spiral spectrum of Bessel–Gaussian beam in non-Kolmogorov turbulence. Optics Communications, 2014, 318(1): 95-99.
[24] J Hu, Z Guo, J Shi, et al. A metasurface-based full-color circular auto-focusing Airy beam transmitter for stable high-speed underwater wireless optical communications. Nature Communications, 2024, 15(1): 2944.
[25] H Yang, Q Yan, Y Zhang, et al. Received probability of perfect optical vortex in absorbent and weak turbulent seawater links. Applied Optics, 2021, 60(35): 10772-10779.
[26] Y Yuan, D Liu, Z Zhou, et al. Optimization of the Probability of Orbital Angular Momentum for Laguerre-Gaussian Beam in Kolmogorov and non-Kolmogorov Turbulence. Optics Express, 2018, 26(17): 21861-21871.
[27] L Gong, Q Zhao, H Zhang, et al. Optical orbital-angular-momentum-multiplexed data transmission under high scattering. Light: Science & Applications, 2019, 8(1): 27.
[28] Y Xu, Y Zhu, Y Zhang. Crosstalk probability of the bandwidth-limited orbital angular momentum mode of Bessel Gaussian beams in marine-atmosphere turbulence. Optics Communications, 2018, 427: 493-496.
[29] A Forbes, L Mkhumbuza, L Feng. Orbital angular momentum lasers. Nature Reviews Physics, 2024, 6(6): 352-364.
[30] S Wang, J Wang, Y Yang, et al. Robust Visible Light Communication via Perfect Vortex Beams. Journal of Lightwave Technology, 2025, 43(16): 7572-7579.
[31] D G Pires, D Tsvetkov, H Barati Sedeh, et al. Stability of optical knots in atmospheric turbulence. Nature Communications, 2025, 16(1): 3001.
[32] H Zhan, L Wang, W Wang. Generative Adversarial Network Based Adaptive Optics Scheme for Vortex Beam in Oceanic Turbulence. Journal of Lightwave Technology, 2022, 40(13): 4129-4135.

[33]   C B Hogge, W L Visinsky. Laser beam probing of jet exhaust turbulence. Applied Optics, 1971, 10(4): 889-892.
[34]   M Bayraktar, B Akın, M B Işık. Propagation of hyperbolic sinusoidal Gaussian beam in jet engine induced turbulence. Optical and Quantum Electronics, 2022, 54(8): 516.